\documentclass{IEEEtran}
\usepackage{graphicx}
\usepackage{amsmath}
\usepackage{amssymb}
\DeclareMathOperator*{\argmin}{arg\,min}

\usepackage{bm}
\usepackage{url}
\usepackage{verbatim}

\newcommand{\set}[1]{\left\{{#1}\right\}}
\newcommand{\eset}[2]{\left\{{#1} : \: {#2}\right\}}
\newcommand{\indic}[1]{1_{\{#1\}}}

\newtheorem{proposition}{Proposition}[section]
\newtheorem{definition}[proposition]{Definition}

\newtheorem{algorithm}[proposition]{Algorithm}

\begin{document}
\title{Regular decomposition of large graphs and other structures: scalability and robustness towards missing data} 
\author{\IEEEauthorblockN{Hannu Reittu and Ilkka Norros}\\
\IEEEauthorblockA{VTT Technical Research Centre of Finland Ltd
\\
hannu.reittu@vtt.fi, ilkka.norros@vtt.fi}\\
\and
\IEEEauthorblockN{F\"{u}l\"{o}p Bazs\'{o}}\\
\IEEEauthorblockA{Wigner Research Centre for Physics, Hungarian Academy of Sciences
\\
bazso.fulop@wigner.mta.hu}}

\maketitle

\begin{abstract}A method for compression of large graphs and  matrices to
a block structure is further developed. Szemer\'edi's regularity lemma is used
as a generic motivation of the significance of stochastic block
models. Another ingredient of the method is Rissanen's minimum
description length principle (MDL). We continue our previous work on the subject, considering cases of missing data and scaling of algorithms to extremely large size of graphs. In this way it would be possible to find out a large scale structure of a huge graphs of certain type using only a tiny part of graph information and obtaining a compact representation of such graphs useful in computations and visualization.
\end{abstract}

\section{Introduction}
So called 'Big Data' is a hot topic in science and applications. Revealing and understanding various relations embedded in such large data sets is of special interest. One good example is the case of semantic relations between words in large corpora of natural language data. Relations can also exist between various data sets, forming large high-order tensors, thus  requiring integration of various data sets. In such a scenario, algorithms based on stochastic block models (SBM, also known as generalized random graph) \cite{decelle} (for a review see e.g. \cite{abbe}) are very attractive solutions instead of simple clustering like k-means. 

It is also natural to look what the abstract mathematics can offer. A strong result, less known among practicing specialists, is so called Szemer\'edi's Regularity Lemma (SRL), \cite{Szeme76} that in way supports SBM-based approach to large data analysis. SRL proves the existence of a SBM-like structure for any large graph and similar objects like hyper-graphs. SRL has been a rich source in pure mathematics --- it appears as an element of important results in many fields. We think that in more practical research, like ours, it is good to be aware of the broad context of SRL as a guiding principle and a source of inspiration.   

The methodology of Regular Decomposition (RD) has been developed in our previous works \cite{nepusz2008,pehkonenreittu,reittujoensuu,reittubazsonorros}. In \cite{reittubazsonorros} we analyzed RD using information theory and Rissanen's Minimum Description Length Principle (MDL) \cite{grunwald}. Our theoretical description is partly overlapping with SBM literature using the apparatus of statistical physics, e.g., extensive works by T.P.\ Peixoto. In particular, the use of MDL to find the number of blocks has been studied earlier by Peixoto \cite {peixoto13}. 

We focus on the case of very large and dense simple graphs. Other cases like rectangular matrices with real entries can be treated in very similar manner using the approach described in \cite{reittubazsonorros}. We wish to demonstrate that large scale block structures of such graphs can be learned from a bounded size, usually quite small, samples. The block structure found from a sample can then be extended to the rest of the graph in just a linear time w.r.t.\ the number of nodes. This means that a large-scale structure of graphs can be found from a limited sample without extensive use of adjacency relations. Only a linear number of links is needed, although the graph is dense with a quadratic number of links. Such a structure is probably enough for many applications like those involving counting of small subgraphs. It also helps in finding a comprehensive overall understanding of graphs when the graph size is far too large for complete mapping and visualization. The graph data can also be distributed, and the process of labeling can then be distributed extremely efficiently since the classifier is a simple object. We also introduce a new version of Regular Decomposition for simple graphs that tolerates missing data. Such an algorithm helps its use in the often encountered situation that part of the link data is missing. 

Our future work will be dedicated to the case of sparse graphs, which is the most important in big data. Here we could merge well-clustered local subgraphs into super-nodes and study graphs of such super-nodes. Provided that we obtain dense graphs, regular decomposition could be used efficiently. SRL has a formulation for the sparse case that can be used as a starting point, along with other ideas from graph theory and practices, to create RD algorithms that are efficient with sparse data. 

\section{Regular decomposition of graphs and matrices}

SRL states, roughly speaking, that the nodes of any large enough graph can be partitioned into a bounded number, $k$, of equally sized 'clusters', and one small exception set, in such a way that most pairs of clusters look like random bipartite graphs with independent links, with link probability equal to the link density. SRL is most significant for large and dense graphs. However, a similar result is extended also to many other cases and structures, see the Refs.\ in \cite{fox2017} with a constant flow of significant new results. 

In RD, we simply replace random-like bipartite graphs by a set of truly random bipartite graphs and use it as a modeling space in the MDL theory. The next step is to find an optimal model that explains the graph in a most economic way using MDL \cite{reittubazsonorros}. In the case of a  matrix with non-negative entries, we replace random graph models with a kind of bipartite Poissonian block models: a matrix element $a_{i,j}$ between row $i$ and column $j$ is thought to correspond to a random multi-link between nodes $i$ and $j$ and the number of links is distributed as a Poisson random variable with mean $a_{i,j}$. The bipartition is formed by the sets of rows and columns, respectively. Then the resulting RD is very similar to the case of binary graphs. This approach allows also the analysis of several inter-related datasets, for instance, using corresponding tensor models. RD also promises extreme scalability as well as tolerance against noise and missing data. 

In RD, the code for a simple graph $G=(V,E)$ with respect to a
partition $\xi=\{A_1,\ldots,A_k\}$ of $V$ has a length at most
(and, for large graphs, typically close to)
\begin{align}
\label{bmcodexi}
\nonumber
L(G|\xi)&=L_1(G|\xi)+L_2(G|\xi)+L_3(G|\xi)+\\
\nonumber
&+L_4(G|\xi)+L_5(G|\xi),\\
\nonumber
L_1(G|\xi)&=\sum_{i=1}^kl^*(|A_i|),\\
L_2(G|\xi)&=\sum_{i=1}^kl^*\left(\binom{|A_i|}{2}d(A_i)\right)+\\
\nonumber
&+\sum_{i<j}l^*\left(|A_i||A_j|d(A_i,A_j)\right),\\
\nonumber
L_3(G|\xi)&=|V|H(\xi),\\
\nonumber
L_4(G|\xi)&=\sum_{i=1}^k\binom{|A_i|}{2}H(d(A_i)),\\
\nonumber
L_5(G|\xi)&=\sum_{i<j}|A_i||A_j|H(d(A_i,A_j)),
\end{align}
where $l^*(m):=\log(m)+\log\log(m)+\cdots$ is the coding length of an integer $m$ and
\begin{equation*}
H(p)=-p\log p -(1-p)\log(1-p)
\end{equation*}
is the entropy of the Bernoulli$(p)$ distribution. A corresponding  representation exists also for the coding length of a matrix, interpreted as an average of a Poissonian block model, see details in \cite{reittubazsonorros}.

\begin{figure}
\centering
\includegraphics[width=9cm,trim=0 100 0 0]{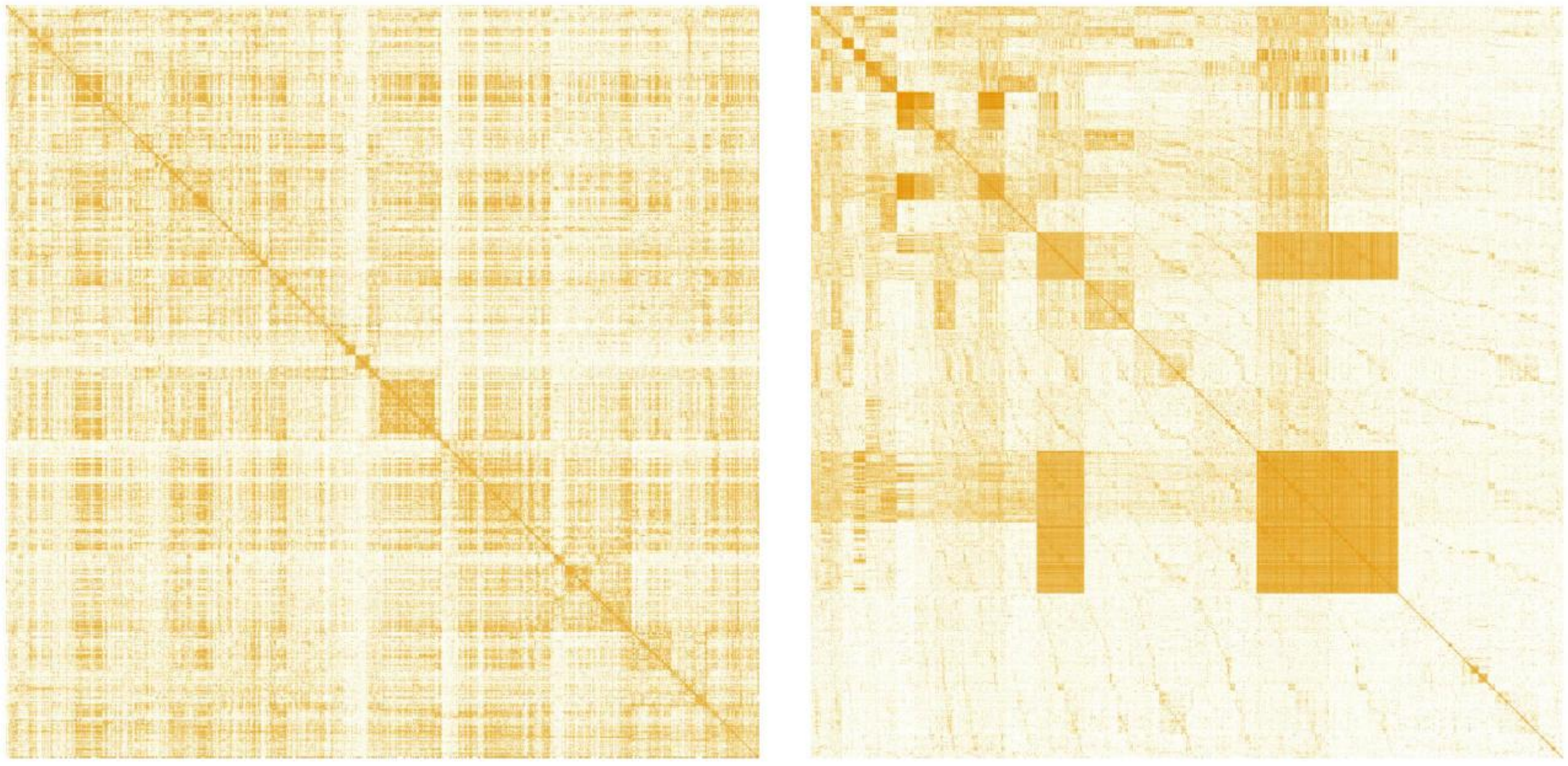}
\caption{Left: similarity matrix between 4088 sentences from the book "Origin of Species" by Charles Darwin; Right: the same matrix where sentences are reordered in 20 regular group revealing a characteristic chessboard structure}
\label{origiregular}   
\end{figure}

Figure \ref{origiregular} shows is a typical  result with regular decomposition of a symmetric real matrix. The regular groups are seen as almost flat rectangular blocks that are revealed by a permutation of rows and columns of the left hand side matrix. For a non-symmetric rectangular matrix, the visual effect of RD is similar: a ``chessboard'' that is irregular in the sizes and shapes of the rectangles.

\subsection{Matrix formulation of RD algorithm for graphs}
Next we present algorithms that we have used in actual
computations of regular decompositions of graph and matrix data \cite{reittubazsonorros}. These
are written corresponding to \eqref{bmcodexi} (except that $L_1(G|\xi)$ is neglected, being insignificant even in the rigorous analysis \cite{reittubazsonorros}). The code lengths $L_4$ and
$L_5$ have a usual interpretation as the minus log-likelihood of a graph
corresponding to a stochastic block model. The version with missing data, presented at the end, is a new contribution. These algorithms use matrix algebra and are supposed to work with moderate size data. In the next section, we propose a sampling-based algorithm that allows the scaling of RD for big data.

A partition of a set of $n$ nodes into $k$
  nonempty sets, called the blocks,  can be described by an $n\times k$ binary matrix such that all row sums equal to one and none of the column sums equals
  zero. The latter condition indicates that the sets of the partition are non-empty. Denote the space of all such matrices by ${\cal{R}}_k$ and
  the members of this set as $R\in{\cal{R}}_k$.

\begin{definition}
For a given graph $G$ with $n$ nodes, adjacency matrix $A$ and a
partition matrix $R\in{\cal{R}}_k$, denote
\begin{equation*}
P_1(R):= R^TAR,
\end{equation*}
where $\cdot^T$ stands for matrix transpose. Block sizes are the column sums of $R$ and are denoted
as 
\begin{equation*}
n_\alpha:=(R^TR)_{\alpha,\alpha},\quad 1\leq\alpha\leq k.
\end{equation*}
The number of links within each block and between block pairs are denoted and computed as
\begin{equation*}
e_{\alpha,\beta}(R)= (1-\frac{1}{2}\delta_{\alpha,\beta})(P_1(R))_{\alpha,\beta}.
\end{equation*}

Finally, define $k\times k$ link-density matrix
\begin{align*}
(P(R))_{\alpha,\alpha}&:=\indic{n_\alpha>1}\frac{e_{\alpha,\beta}(R)}{{n_\alpha\choose 2}},\\
(P(R))_{\alpha,\beta}&:=\frac{e_{\alpha,\beta}(R)}{n_\alpha n_\beta},\ \alpha\not=\beta.\\
\alpha,\beta&\in\{1,2,\cdots,k\}
\end{align*}
\end{definition}

Then the  length of the code that uniquely describes the graph corresponding to $A$, using the model $(R,P)$, appears to be:

\begin{definition}
\label{algcodelendef}
\begin{align*} 
l_k(G(A)\mid R\in{\cal R}_k):=& \sum_{1\leq i< j\leq k}n_in_j H((P(R))_{i,j})+\\
&+\sum_{1\leq i\leq k}{n_i\choose 2}H((P(R))_{i,i}) + l_k(R),
\end{align*}
where 
$$
l_k(R)= \sum_{1\leq i\leq k}n_iH(n_i/n)+ \sum_{1\leq i\leq j\leq k}l^*(e_{i,j}(R))
$$ 
presents the code length of the model.
\end{definition}

The two-part MDL program of finding the optimal model, denoted
as $R_{k^*}$, can now be written as:
\begin{eqnarray}
\boxed{(k^*,R_{k^*})
:=\argmin_{1\leq k\leq n}\min_{R\in{\cal R}_k }l_k (G(A)\mid R\in{\cal R}_k)}
\end{eqnarray}
To solve this program approximately, we can use the following greedy algorithm.

\begin{algorithm}
{\bf  {Greedy Two-part MDL}}\\
{\bf Input:} $G=G(A)$ is a simple graph of size $n$.\\
{\bf Output:} $(k^*,R_{k^*}\in {\cal R}_{k^*}), k^*\in\{1,2,\cdots,n\}$, such that the two-part code for $G$ is close to the shortest possible for all possible block models with number of blocks in the range from $1$ to $n$.\\ 
Start: $k=1$, $l^*=\infty$, $R\in{\cal R}_n=\set{I}$, 
$k^*=1$, where $I$ is denotes the $n\times n$ unit matrix.\\
{\bf 1. Find} 
$$
\hat{R}_k(G):=\argmin_{R\in {\cal R}_{k}} (l_k (G\mid R) 
$$
using subroutine {\bf ARGMAX k} (Algorithm \ref{argmaxk}).
\newline
{\bf 2. Compute } $l_k(G)= \lceil l_k(G\mid \hat{R}(G))\rceil+l_k (\hat{R}(G))$\\
{\bf 3.  If} $l_k(G)< l^*$  then $ l^*=l_k(G)$, $R_{k^*}=\hat{R}_k(G)$ , $k^*=k$\\
{\bf 4.} $k=k+1$\\
{\bf 5. If } $k>n$, {\bf Print} $(R_{k^*},k^*)$ and {\bf STOP} the program. \\
{\bf 6.} {\bf GoTo 1}.
\end{algorithm}

\begin{definition}
A mapping $\Phi:{\cal{R}}_k\rightarrow {\cal{R}}_k$ is defined as follows. First, define following matrices element-wisely:
\begin{align*}
(LogP(R))_{\alpha,\beta}&:=\log(P(R)_{\alpha,\beta}),\\ 
(Log[1-P(R)])_{\alpha,\beta}&:=\log[1-P(R)_{\alpha,\beta}],\\
(1-A)_{\alpha,\beta }&:=1-A_{\alpha,\beta},\\
 L(R)&:= -AR(LogP(R))^T-\\
&-(1-A) R Log[1-P(R)],
\end{align*}
where we set by definition, $\log 0:=0$ since it appears in combination $0\log0=0$ in matrix $L$, and $\alpha,\beta\in\{1,2,\cdots,k\}$, and
\begin{equation*}
\beta(i,R):= \inf\{\beta:\beta=\argmin_{1\leq\alpha\leq k}(L(R))_{i,\alpha}\}, 1\leq i\leq n,
\end{equation*}
and finally define the matrix function $\Phi(\cdot)$ element-wisely on argument $R$:
$$
\Phi(R)_{i,\alpha}= \delta_{\alpha,\beta(i,R)}, i\in\{1,2,\cdots,n\}.
$$
\end{definition}

The mapping $\Phi(R)$  defines a sparse optimization of partition $R$, where each node is re-placed to new block independently of all other placements; hence the term 'greedy algorithm'.   

\begin{algorithm}
\label{argmaxk}  
{\bf ARGMAX k}
\newline 
Algorithm for finding regular decomposition for fixed $k$.\\
{\bf Input}: $A$: the adjacency matrix of a graph (an $n\times n$ symmetric binary matrix with zero trace); $N$: an integer (the number of iterations in the search of a global optimum); 
$k$: a positive integer.\\
Start: $m=1$.\\
{\bf 1.}  $i:=0$; generate a uniformly random element $R^i\in\mathcal{R}_k$.\\
{\bf 2.}  If at least one of the column sums of $R^i$ is zero, {\bf GoTo 1}, if not, then compute: 
$$
R^{i+1}:=\Phi(R^i).
$$
{\bf 3.} {\bf If}  $R^{i+1}\neq R^i$, set $i:=i+1$ and {\bf GoTo 2},\\
{\bf If}   $R^{i+1} = R^i$,  {\bf GoTo 4}\\
{\bf 4.} $R(m):=R^i$; $m=m+1$;\\ 
$l(m):= \sum_{i=1}^n \min_{1\leq\alpha\leq k}(L(R(m)))_{i,\alpha}$.\\
{\bf 5.} If $m<N$, {\bf GoTo 1}. \\
{\bf 6.} $M:=\{m: l(m)\leq l(i),i=1,2,...,N\}$; $m^*:=\inf M$. \\
{\bf Output} Solution: $R(m^*)$.
\end{algorithm}
In {\bf ARGMAX k} the outer loop 1-5 runs several ($N$) optimization rounds finding each time a local optima, finally the best local optima is the output. At each optimization round, the inner loop 2-3, improves the initial random partition until a fixed point is reached and no further improvements of the partition are possible. This fixed point is an indication that a local optimum is found.

For very large graphs, the program may not be solvable in the sense
that it is not possible and reasonable to go through all possible
values of $k\in\{1,2,\cdots, n\}$.  One option is to limit the range
of $k$. In case that no minimum is found, then use as an optimal
choice the model found for the largest $k$ within this range. Another
option is to find the first minimum with smallest $k$ and stop. When
the graph is extremely large, it makes sense to use only a randomly
sampled sub-graph as an input --- indeed, when $k^*<<n$, a
large-scale structure can be estimated from a sample, as described in more details in the next section.

\subsection{Regular decomposition with missing link information}

Now assume that part of the link information is lost. We only consider the simplest case when the link information (whether there is a link or not) is lost uniformly over all node pairs. We also exclude possibility of a sparse representation of graphs, when only those node pairs that have links are listed, since this would lead to unresolvable ambiguity in our setting since if some node pair is not in the list it could mean two things: there is no link or the link information is not available.

We just formulate how the matrices used in previous section are modified. The RD algorithms themselves are unaltered. 

Denote by $A$ the adjacency matrix with missing link values (0 or 1) replaced by  $-1$'s. $A$ is symmetric and we also assume that we know its true dimension (the number of nodes). Put $-1$ on diagonal of $A$, just for convenience of further formulas. Define
$$
D:=\frac{A+|A|}{2},
$$
where the absolute value of a matrix is defined element-wise. Thus, $D$ has the same elements as $A$, except that the entries equal to $-1$ are replaced by $0$. Next define a 'mask matrix'
$$
B:=\frac{A-|A|}{2}+1,
$$
where $+1$ means element-wise addition of 1 to all elements of the matrix. That is, $ b_{i,j}=0$ iff $a_{i,j}=-1$, and $b_{i,j}=1$ otherwise. In other words, $B$ is the indicator of known link data in $A$. 
The matrix $P_1$ is now:
$$
P_1(R):=R^TDR.
$$
The number of observed links in a block indexed by $(\alpha,\beta)$ is
$$
e_{\alpha,\beta}=(1-\frac{1}{2}\delta_{\alpha,\beta})(P_1(R))_{\alpha,\beta}.
$$
We also need the following matrix, whose elements present the number of node pairs in block $(\alpha,\beta)$ that contain link data:
\begin{align*}
(N)_{\alpha,\beta}= n_{\alpha,\beta}&:= (1-\frac{1}{2}\delta_{\alpha,\beta})\sum_{i,j}r_{i,\alpha}r_{j,\beta}b_{i,j}\\
&=(1-\frac{1}{2}\delta_{\alpha,\beta})(R^TBR)_{\alpha,\beta}.
\end{align*}
The $P$-matrix is defined as the link densities in the observed data:
\begin{align*}
(P(R))_{\alpha, \beta}:=1_{\{n_{\alpha,\beta}>0\}}\frac{e_{\alpha,\beta}}{n_{\alpha,\beta}}.
\end {align*}
As stated above, we assume that the matrix elements of $P$ be close to the corresponding actual link densities in the case that we would possess complete link data. This also assumes that the graph and the blocks must have large size, which is not a big constraint since RD is aimed at the analysis of large graphs.

The cost function in Definition \ref{algcodelendef} remains unaltered, although now the $P$-matrix elements are only estimates based on observed data and not the exact one. The block sizes are true ones that are known since we assume that we know the actual number of nodes in the graph. The interpretation is that the contributions of missing data entries in a block are replaced by the average contribution found from the observed data of that block. It should be noted that similar scaling must be done in the RD algorithms that find the optimum of the cost function. E.g., $A$ should be replaced with $D$, and summing over elements of $D$ should be rescaled so that the contribution is proprtional to the true size of the corresponding block.

\section{Regular decomposition of large graphs using sparse sampling}

\subsection{Sampling scheme and experimental testing}

Assume that a very large graph is generated from a SBM. Denote the blocks or node clusters by $V_1,V_2,\cdots, V_k$ and the link probability matrix by $P$, that is, a $k\times k$ symmetric matrix with entries in $(0,1)$. We also assume that no two rows are identical, to avoid redundant structure. Then we generate links between blocks and within blocks using $P$ as independent link probabilities. The size of the graph $G$ is assumed large and denoted by $N$. 

Now take a uniform random sample of size $n$ of the nodes. Denoted the graph induced by sampling as $G_n$. The relative sizes of blocks are denoted as $r_i:=\frac{|V_i|}{N}$. The probability of sampling a node from block $i$ is $r_i$.  We want to test whether the block structure of $G$ can be reconstructed from a small sample, the size of which does not depend on $N$, using a classifier with time complexity $\sim N$. We believe that the answer is "yes". 

For simplicity, assume that we know the block structure of $G_n$ by having run an ideally working RD-algorithm. This is not too far from reality, although some of the nodes may in reality be misclassified. However, there is not a big difference, if most of the nodes are correctly classified and $n$ is not very small. The real graphs, however, are not generated from any SBMs, and this is a more serious drawback. In the latter case we would like to show that it is not necessary to run RD on $G$, rather it is sufficient to run RD on $G_n$ and just classify the rest of the nodes based on that structure. The linkage between the SBM case and the case of real data is not rigorous and needs further work.

In $G_n$, we denote the blocks as $\hat{V}_1,\hat{V}_2,\cdots, \hat{V}_k$, and the empirical link densities between these blocks as $\hat{d}_{i,j}$. The relative sizes of the blocks as well as the link densities deviate in $G_n$ from their counterparts in $G$. 
Now assume that we take a node $i\in V_\beta$ outside $G_n$ together with its links to $G_n$. What is the probability that the RD classifier places this node into $\hat{V}_\beta$? 

Denote by $e_\alpha(v)$ the number of links from $v$ to $\hat{V}_\alpha$, $n_\alpha:=|\hat{V}_\alpha|$. The RD classifier based on $G_n$ is the following program:
\begin{align*}
C_\alpha(v|\hat{\xi},\hat{d},k)&:=\sum_{j=1}^k\big[-e_j(v)\log\hat{d}_{j,\alpha}\\
&\hspace*{15mm}-(n_j-e_j(v))\log(1-\hat{d}_{j,\alpha})\big],\\
\alpha^*&=\argmin_\alpha(C_\alpha(v|\hat{\xi},\hat{d},k)),
\end{align*}
where $\hat{\xi}=\set{\hat{V}_1,\hat{V}_2,\cdots, \hat{V}_k}$ and $\hat{d}$ stands for the $k\times k$ matrix with matrix elements $\hat{d}_{i,j}$.
To compute the optimal block for node $i$ we need to compute and check a list of $k$ numbers as well as compute the densities  $e_\alpha(i)$. The computation time is upper-bounded by $\Theta(k n)$. For a size $n$ that is independent on $N$, the computation takes only a constant time. By this we mean that it is enough to have some constant $n$ regardless how large the graph is. Such an $n$ depends on the relative sizes of blocks and on $k$. This is because it is obviously necessary to have enough samples from each block to be able to obtain good estimates of link densities and if some blocks are very small, more samples are needed.

We made experiments with $k=20$ and with equal relative sizes of blocks. The $P$-matrix elements were i.i.d.\  Uniform$(0,1)$ r.v.'s. In this case, already $n>200$ is such that no classification error was found in a long series of experiments with repetitions of random samples and classification instances. Of course, errors are always possible, but they become insignificant already when each block has about 20 members. A similar situation is met with non-equal block sizes, but then the small blocks dictate the required $n$. When the smallest block has more that a couple of dozen nodes, the errors become unnoticeable in experiments.

We conjecture that if such a block structure or close to it exists for a large graph, it can be found out using only a very small subgraph and a small number of links per node in order to place every node into right blocks with only very few errors. This would give a coarse view on the large graph useful for getting an overall idea of link distribution without actually retrieving  any link information besides labeling of its nodes. For instance, if we have a large node set, then the obtained labeling of nodes and the revealed block densities would allow computation of, say, the number of triangles or other small subgraphs, or the link density inside a set. It would be possible to use this scheme in case of dynamical graph, to label  new nodes and monitor the evolution of its large scale structure. We can also adapt our RD method to the case of multi-layer networks and tensors and etc. A similar sampling scheme would also be desirable and probably doable. On the other hand, large sparse networks need principally new solutions and development  of a kind of sparse version of RD. These are major directions of our future work. 
  
\begin{figure}[ht]
\centering
\includegraphics[width=5cm]{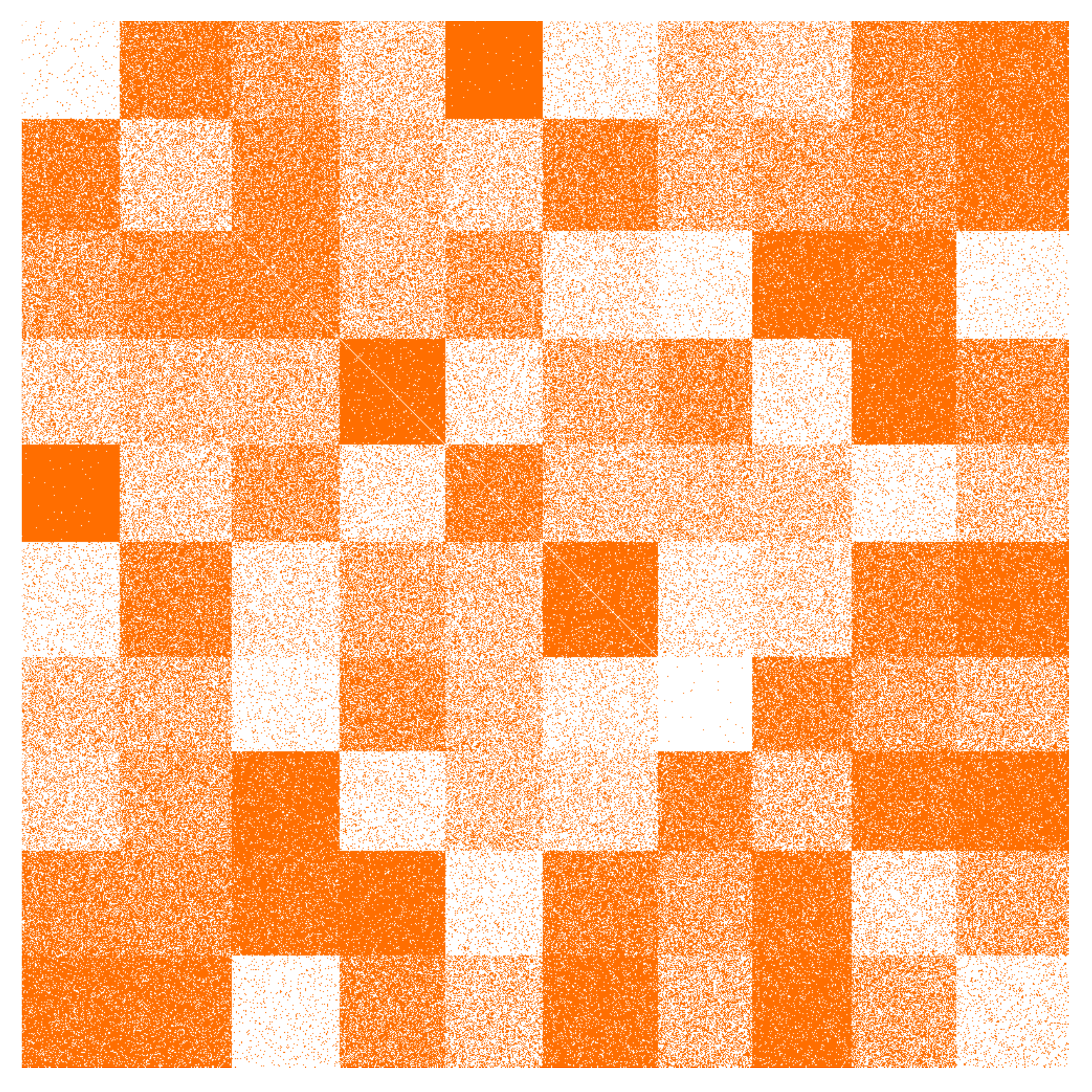}
\caption{Adjacency matrix of a random graph with $k$=10, and equal size blocks and $1000$ nodes. }
\label{sample1000}       
\end{figure}

\begin{figure}[ht]
\centering
\includegraphics[width=5cm]{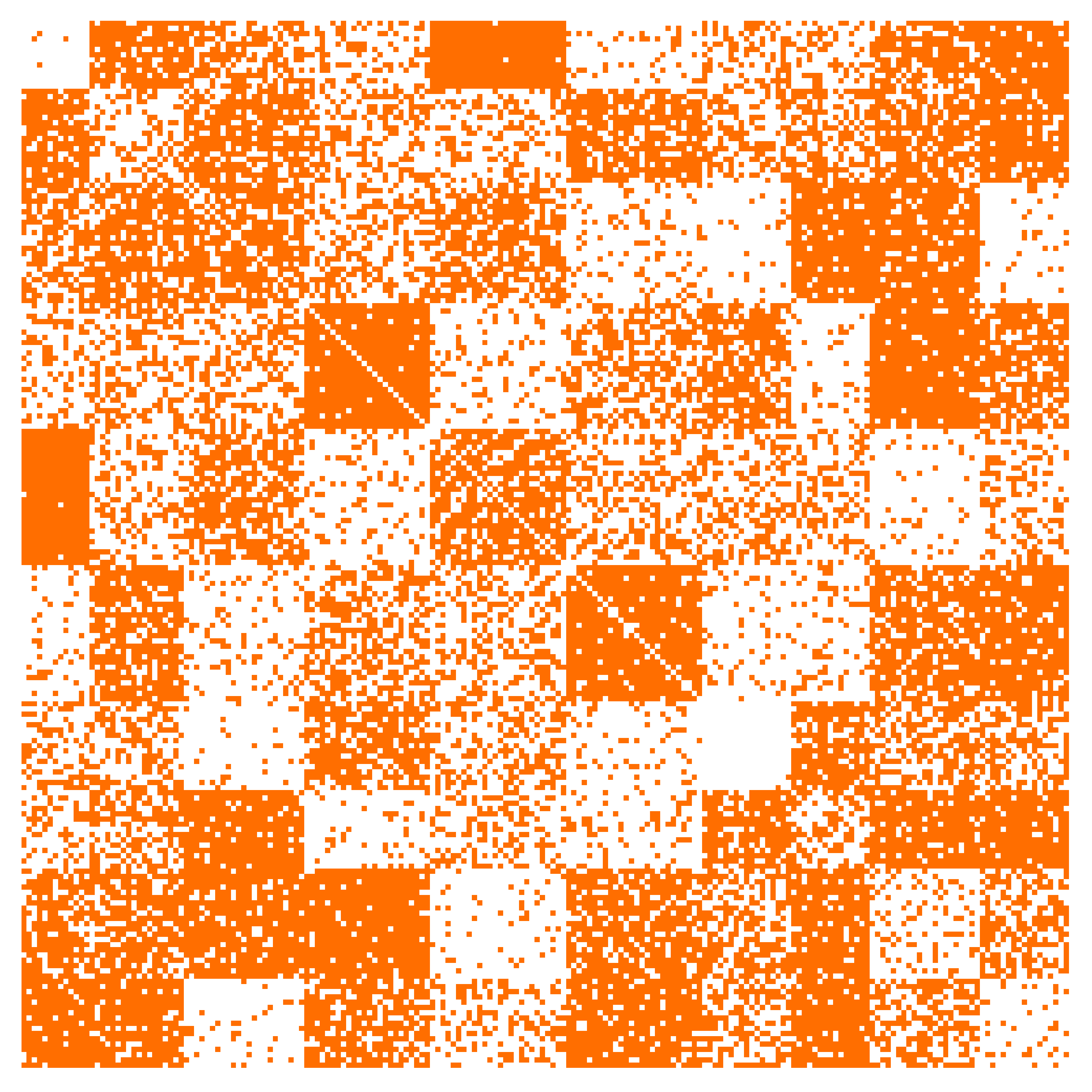}
\caption{Adjacency matrix of a random graph with $k$=10, and equal size blocks, generated from the same model as in previous picture but with only $200$ nodes.  }
\label{sample200}       
\end{figure}

\begin{figure}[ht]
\centering

\includegraphics[width=7cm]{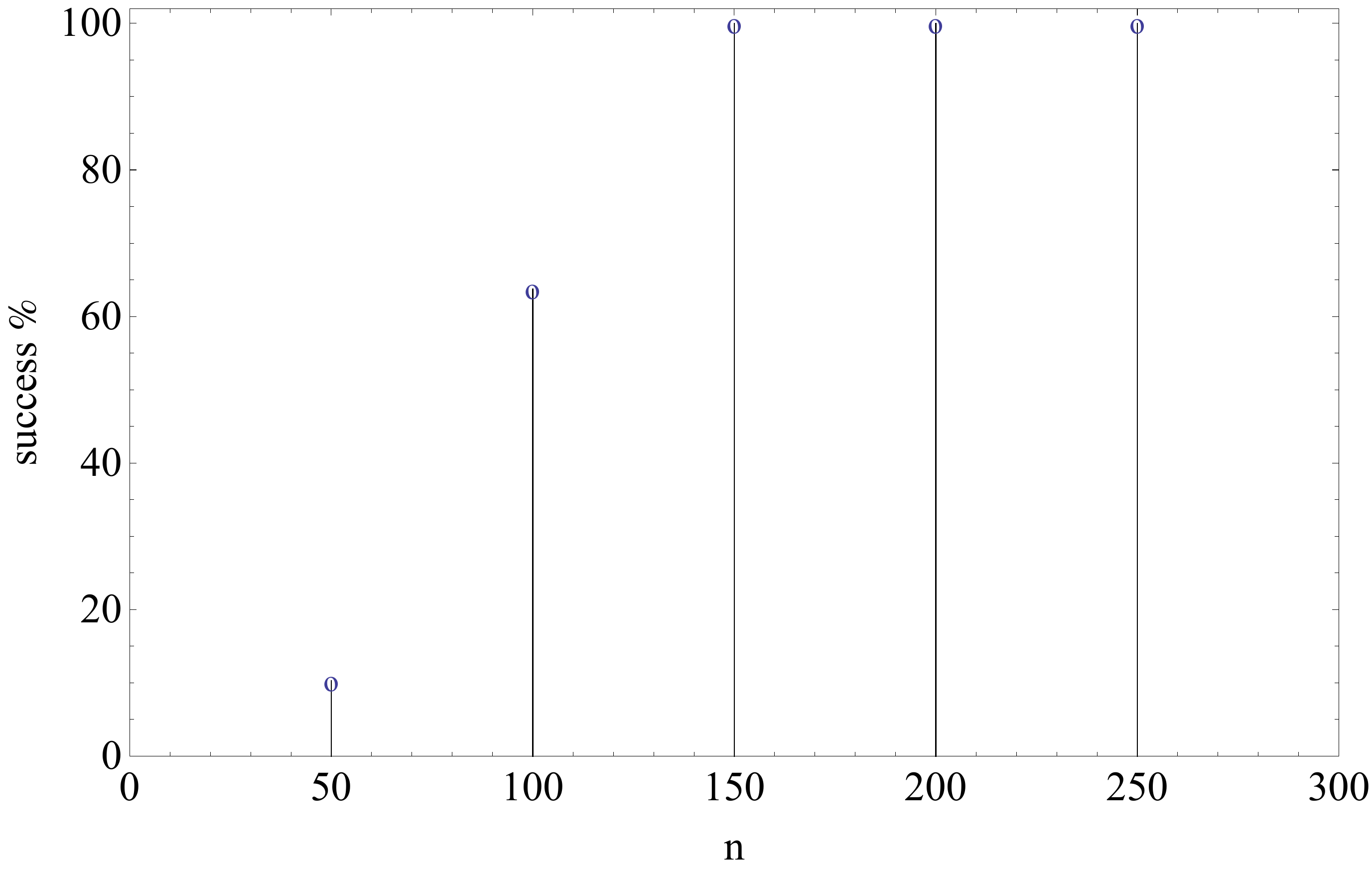}

\caption{Classification success percentile as function of $n$ using several repetitions of experiments and a 1000 classification instances for each. Already a sample with 200 nodes generates a model with almost a perfect classifier}
\label{success}       
\end{figure}

\begin{center} 
\begin{figure}[!tbp]
  \centering
  \begin{minipage}[b]{0.4\textwidth}
\centering
    \includegraphics[width=4.5cm]{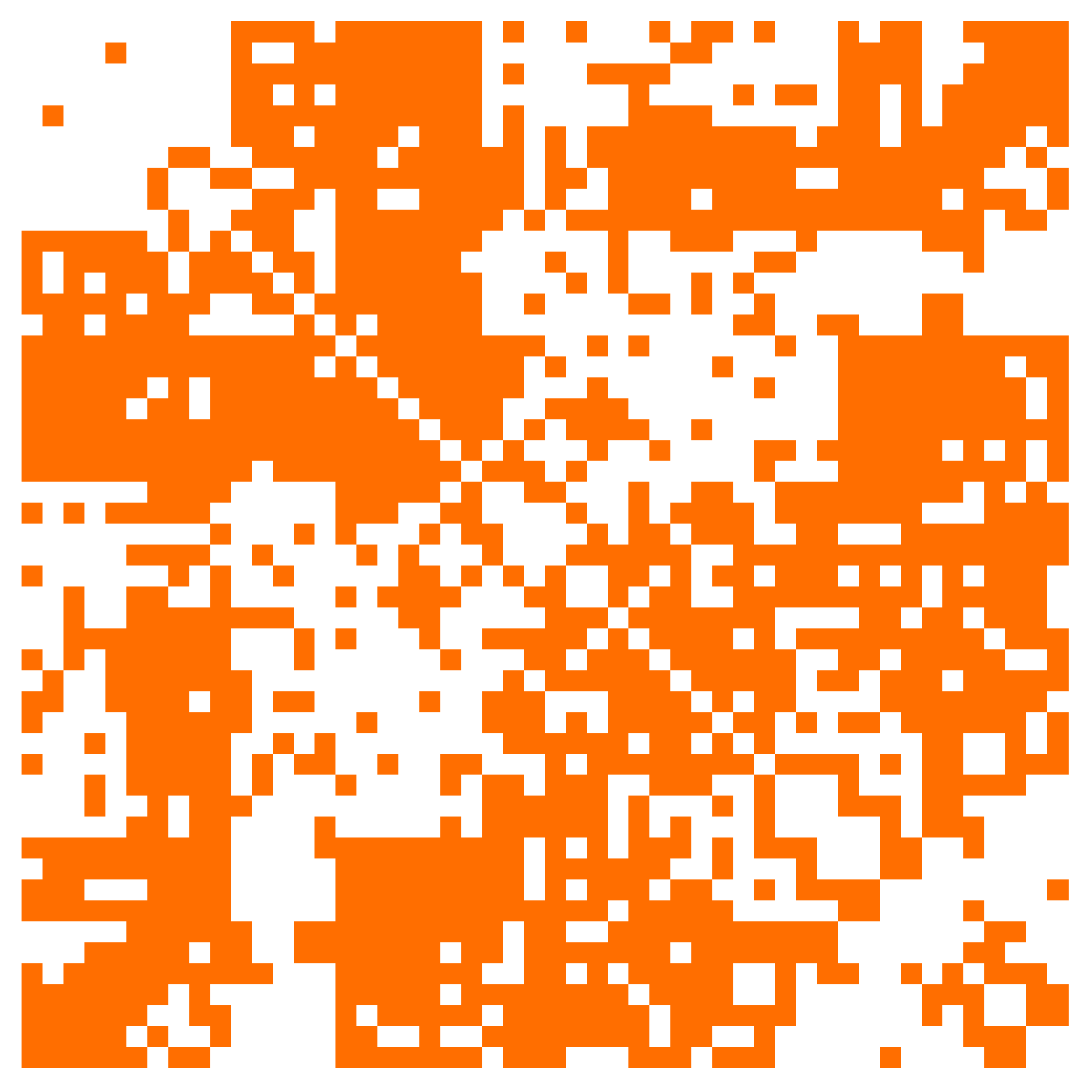}
    \caption{A sample graph with 50 nodes that is insufficient to create a successful classifier - the result is similar to completely random classification.\newline}
  \end{minipage}
  \hfill
  \begin{minipage}[b]{0.4\textwidth}
\centering
    \includegraphics[width=4.5cm]{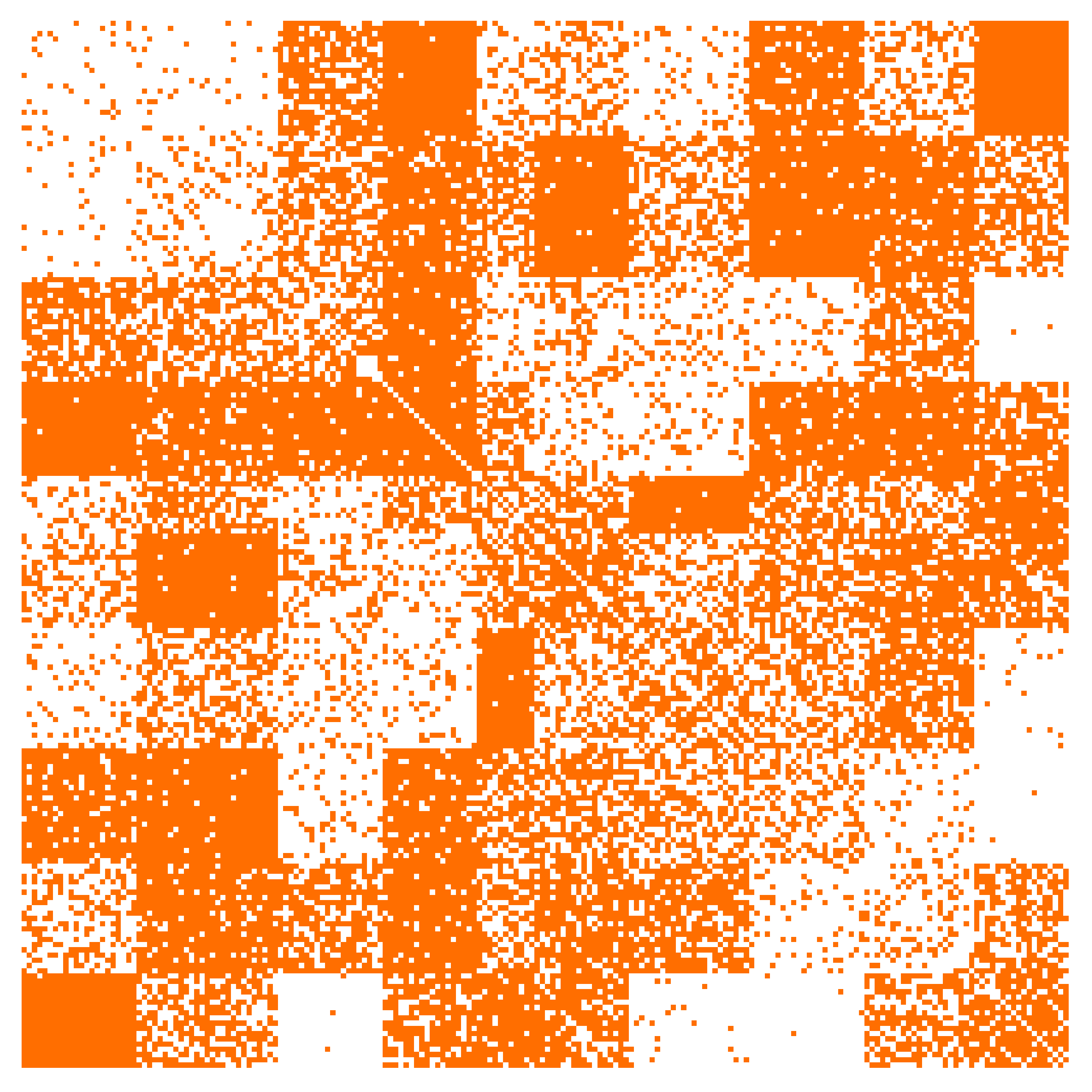}
    \caption{200 node sample, from the same model as above,  that generates almost a perfect classifier - no errors detected in experiments. }
  \end{minipage}
\end{figure}
\end{center}
\begin{figure}[ht]
\centering
\includegraphics[width=8.7cm,trim=0 80 0 0]{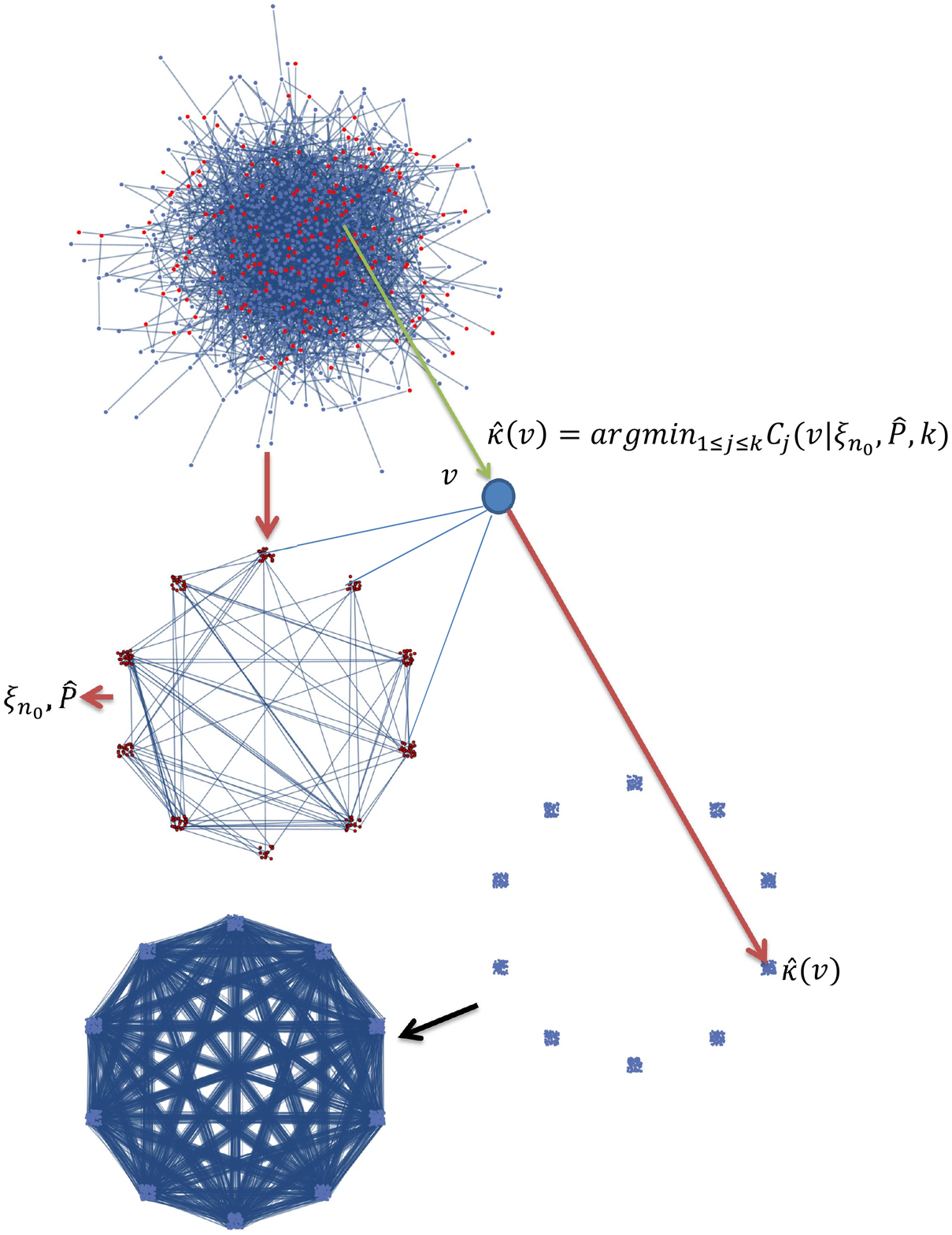}
\caption{A scheme of RD for a huge graph shown at the top, in reality we assume much larger graph than what is shown in the picture. First a moderate size sample of $n_0$ nodes and induced subgraph is formed. RD is applied to this small graph, groups $\hat{\xi}_{n_0}$ and matrix $\hat{P}$ are found, shown as the graph in the middle. The sequentially any node $v$ from the big graph can be classified to a regular group using just counts of links of this chosen node to regular groups of the sample graph. This classification requires a constant number of computations, upper bounded by $k^2 n_0$, with elementary functions. As a result nodes of the whole graph can be classified in just linear time with respect to number of nodes. After the classification is done (shown in the ring shape structure at the right side) the RD of the large graph is done simply by retrieving link information. The result is shown in the lower graph. }
\label{sampling}       
\end{figure}

\subsection{Mathematical framework for rigorous analysis}

In our future work we aim at rigorous analysis of errors in the sampling scheme described in the previous section. Here we give some more convenient formulations that we shall use in our future work. They also clarify the character of mathematical objects we are dealing with.

Obviously, we can generate a graph process $G_n(V_n,E_n,\xi_n)$ of
increasing realizations of Blockmodel$(r,P)$ by starting with a single
vertex and adding at every step $n+1$ one new vertex $v_{n+1}$, its label $\kappa(v_{n+1})$ and its edges to $V_n$.

Now assume that the labellings are hidden after some $n_0$, and only
the edges are observed. Instead, each label $\kappa(v_n)$, $n>n_0$, is
estimated based on the edge set $e(v_n,V_0)$ and the empirical model
Blockmodel$(\hat{r},\hat{P})$. In this way, a sequence of
classifications $\hat{\xi}_n$, $n>n_0$, is generated. How different
are $(\xi_n)$ and $(\hat{\xi}_n)$?

Note that, given $(V_{n_0},E_{n_0},\xi_{n_0})$, the set-valued process
$\eset{w\in V_0}{(v_n,w)\in E_n}$ is i.i.d., and therefore also the
process $\hat{\kappa}(v_n)$ is i.i.d.\ for $n>n_0$.

The maximum likelihood classifier $\hat{\kappa}(\cdot)$ has the form
\begin{align}
\label{kappahatdef}
\hat{\kappa}(v)=\argmin_jC_j(v| \xi_{n_0},\hat{P},k) .
\end{align}
Denote

\begin{equation*}
q_i(v)=\frac{|e(v,\hat{U}_i)|}{|\hat{U}_i|}.
\end{equation*}
Denote the Kullback-Leibler divergence from Bernoulli($p$) to Bernoulli($q$) as 
\begin{equation*}
I(q:p)=\frac{q}{p}\log\frac{q}{p}+\frac{1-q}{1-p}\log\frac{1-q}{1-p}.
\end{equation*}
Adding and subtracting $H(q_i(v))$ to each summand in
\eqref{kappahatdef}, the expression to be minimized can be written as
\begin{align*}
&\sum_{i=1}^k|\hat{U}_i|\left(I(q_i(v):\hat{p}_{ji})+H(q_i(v))\right)\\
&=|V_0|\left[\sum_{i=1}^n\hat{r}_iI(q_i(v):\hat{p}_{ji})
  +\sum_{i=1}^n\hat{r}_iH(q_i(v))\right].
\end{align*}
Because the second sum does not depend on $j$, the definition
\eqref{kappahatdef} obtains the more intuitive form
\begin{equation}
\label{kappahatdef2}
\hat{\kappa}(v)=\argmin_j\sum_{i=1}^k\hat{r}_iI(q_i(v):\hat{p}_{ji}).
\end{equation}
Define also the ``ideal $(V_{n_0},E_{n_0},\xi_{n_0})$ based classifier''
\begin{equation*}
\hat{\kappa}^*(v)=\argmin_j\sum_{i=1}^kr_iI(q_i(v):p_{ji}).
\end{equation*}
Note that we don't always have $\kappa(v)=\kappa^*(v)$.

In future work we try to find proofs and conditions for the described sampling scheme assuming also partially lost data and without the assumption that graph is generated using a SBM. Here the 'Sampling lemma' 2.3 from \cite{fox2017} can be handy, proving that in general setting the link densities can be found from small samples. Even more intriguing is the question of using RD for sparse graphs and finding some kind of analog of RD-approach in this situation.


\bigskip
{\bf Acknowledgment.} The work of the Finnish authors was supported by Academy of Finland project 294763 (Stomograph).


\end{document}